\documentclass[11pt,twoside]{article}

\usepackage{asp2004}
\usepackage{epsf}
\usepackage{psfig}
\usepackage{lscape}
\usepackage{graphicx}

\newcommand{\msun}{\mbox{$M_{\odot}$}}
\newcommand{\lsun}{\mbox{$L_{\odot}$}}
\newcommand{\zsun}{\mbox{$Z_{\odot}$}}
\newcommand{\Zsun}{\mbox{$Z_{\odot}$}}
\newcommand{\Teff}{\mbox{$T_{\rm eff}$}}
\newcommand{\vinf}{\mbox{$v_{\infty}$}}
\newcommand{\mdot}{\mbox{$\dot{M}$}}
\newcommand{\msunyr}{\mbox{$M_{\odot} {\rm yr}^{-1}$}}
\newcommand{\ratio}{\mbox{$v_{\infty}$/$v_{\rm esc}$}}

\begin{document}
\title{Massive star feedback -- from the first stars to the present}   
%
\author{Jorick S. Vink$^{1,2}$}  
\affil{$^1$Imperial College London, Blackett Lab, Prince Consort Rd, UK\\
       $^2$Keele University, Astrophysics, Lennard-Jones Lab, ST5 5BG, UK} 
\begin{abstract} 
The amount of mass loss is of fundamental importance to the lives and deaths of very 
massive stars, their input of chemical elements and momentum into the interstellar and intergalactic media, as well as 
their emitted ionizing radiation.
I review mass-loss predictions for hot massive stars as a function 
of metal content for groups of OB stars, Luminous Blue Variables, and Wolf-Rayet stars. 
Although it is found that the predicted mass-loss rates drop steeply with decreasing metal 
content ($\mdot$ $\sim$ $Z^{0.7-0.85}$), I 
highlight two pieces of physics that are often overlooked: (i) mass-loss predictions 
for massive stars approaching the Eddington limit, and for (ii) stars that have enriched their own atmospheres with 
primary elements such as carbon. Both of these effects may significantly boost the mass-loss rates 
of the first stars -- relevant for the reionization of the Universe, and a potential pre-enrichment of the intergalactic 
medium -- prior to the first supernova explosions.
\keywords{stars: early-type -- stars: mass loss -- stars: winds, outflows -- stars: evolution -- stars: Wolf-Rayet -- cosmology: early universe}
\end{abstract}


\section{Introduction: stellar cosmology}   
Over the last couple of years, we have witnessed a large increase in the 
interest in hot luminous stars. The reasons for this are manifold. Some of the eye-catchers have 
been the identification of long-duration gamma-ray bursts (GRBs) with supernova 
explosions (Galama et al. 1998, Hjorth et al. 2003) as well as massive star formation 
at high redshift (Bloom et al. 2002). In addition, the impressive numerical simulations 
regarding the formation of the first stars at zero 
metallicity (Population {\sc iii}) -- widely believed to be very massive stars (VMS) with 
$M$ $\sim$ 100\msun\ (Abel et al. 2002, Bromm et al. 1999) -- 
have triggered increased attention into the workings of massive stars at low metallicity ($Z$).

Aside from these relatively recent developments, hot stars with their strong winds have 
continuously been at the forefront of astrophysical research. This is largely due to their 
roles in shaping their environments, by releasing large amounts of mechanical energy via their winds and 
supernovae, as well as via their vast quantities of ionizing radiation. 

It is the {\it amount} of mass loss that lies at the heart of massive star feedback, as it is 
this very quantity that determines: (i) the evolution of massive stars towards their final 
phases, including their ultimate fate, (ii) the density structure of the 
atmosphere, which, in turn sets the terminal wind velocity, and the wind 
energy release into the interstellar medium (ISM), and finally (iii) the hardness of the  
ionizing radiation field (see Gabler et al. 1989). 

Now that the high-redshift Universe has become increasingly more accessible, both with wonderful 
detections of the integrated light of massive stars in star-bursting galaxies 
at $z$ $>$ 6 (e.g. Bunker et al. 2004), as well as with fingerprints of {\it individual} massive stars via their 
associated GRBs (for instance with {\sc swift}), it is critical to understand the nature of massive stars 
and their feedback as a function of metal content. 

It has been known for many years, that mass-loss uncertainties as small as  
a factor of two are able to completely change massive star model output (Meynet et al. 1994), but 
the uncertainties relating to mass loss at very low ($Z/\Zsun$ $\sim$ $10^{-2}$) and 
extremely low ($Z/\Zsun$ $\la$ $10^{-3}$) metallicity are highly uncertain and could easily amount to factors 
of thousands. 

Properly understanding and accurately {\it predicting} the mass-loss rate 
as a function of metal content is therefore of prime importance for reliably assessing the 
direct role of massive star winds in pre-enriching the intergalactic medium (IGM) at early epochs, their 
intricate role in determining the evolution of Population {\sc iii} stars to their final explosions, 
and the amount of ionizing radiation that 
may have significantly contributed to the reionization of the Universe. 

For these reasons, mass-loss predictions as a function of metallicity are at the 
heart of ``Stellar Cosmology''.
Since individual massive star winds are not observable at metallicities below $Z/\zsun$ $\sim$ $10^{-1.5}$, we must rely on 
a comprehensive modelling procedure to allow for a quantitative comparison with observational analyses of stars in 
the local Universe, as well as a robust method for extrapolating these predictions into the extremely low $Z$ ($Z/\zsun$ $\la$ $-3$) 
domain, where we can no longer observe stellar winds {\it directly}. 

In the next section, I describe the physical process of radiation pressure on spectral lines, 
widely accepted to drive massive star winds, and continue to introduce  
two approaches currently in use for predicting mass-loss rates of massive stars, 
the CAK and Monte Carlo approach, with their pros and cons. 
In section 3, the successes and remaining challenges of these predictions are described for a 
range of massive objects comprising 
OB supergiants, Luminous Blue Variables (LBVs), and Wolf-Rayet (WR) stars, with a prime focus 
on their metallicity dependence. I finish with a discussion highlighting cosmological 
implications of these initial results. 

\section{Two methods to predict hot-star mass-loss rates} 
Light carries momentum, and it has been known since the 1920s (e.g. Milne 1926) 
that radiation pressure on spectral lines is able to eject atoms from stars. Nonetheless, 
it was not until the late 1960s, when sensitive UV mass-loss diagnostics of O star winds 
became available, that radiation pressure on spectral lines came back into the foreground of 
astrophysical research.

\subsection{The line-driven wind theory and the CAK approach}
Lucy \& Solomon (1970) were the first to realize that radiation pressure on metal 
lines, in combination with sharing this momentum with the abundant 
hydrogen plasma, could drive a continuous outflow from hot luminous stars. The necessary condition 
for initiating a wind is that the radiative acceleration $g_{\rm rad}$ exceeds gravity, as can be  
seen in the momentum equation:

\begin{equation}
v \frac{dv}{dr}~=~- \frac{G M}{R^2}~+~g_{\rm rad}
\label{eq_motion}
\end{equation}
where the gas pressure term is neglected. 
For a more comprehensive review on stellar wind theory, and 
in particular how the various relevant forces determine a star's mass-loss rate and 
terminal wind velocity, I refer the reader 
to Lamers \& Cassinelli (1999). 

The challenge lies primarily in accurately 
predicting the $g_{\rm rad}$ term in the equation of motion. 
For free electrons this is simply the electron opacity, $\kappa_{\rm e}$, times the flux:

\begin{equation}
g^{\rm elec}_{\rm rad}~=~\frac{\kappa_{\rm e} L}{4 \pi r^2 c}
\label{eq_elec}
\end{equation}
However line scattering turns out to be the more dominant contributor to the total radiative force. 
The reason for this is that although photons and matter are allowed to interact 
only at very specific frequencies, the two can be made to ``resonate'' at a wide range 
of positions in the stellar wind due to the Doppler effect (see reviews by Abbott 1984, Owocki 1994).
Historically, the total radiative acceleration in hot star atmospheres is effectively the force due to free electrons 
{\it multiplied} by a factor $M$, the line Force Multiplier introduced by Castor et al. (1975; CAK), 
for the contribution of strong and weak spectral lines. 
$M(r,t,dv/dr)$ is a function of distance in the wind $r$, optical depth $t$, 
and velocity gradient, $dv/dr$. 
In this formulation the radiative acceleration depends itself on the Newtonian acceleration 
from the inertia term and the equation of motion (Eq.~\ref{eq_motion}) becomes non-linear (e.g. Lucy 1998).
CAK computed this Force Multiplier for an ensemble of carbon {\sc iii} lines in Local Thermodynamic 
Equilibrium (LTE) and found their $M(t)$ to be a linear function of optical depth, so that they were able to fit
this function with two parameters 

\begin{equation}
M(t)~=~k~t^{-\alpha}
\label{eq_cak}
\end{equation}
where $k$ and $\alpha$ are the famous CAK force multiplier parameters, $k$ being a measure for the 
number of lines, and $\alpha$ representing the number of strong to weak lines. Several interesting
papers have appeared in the last three decades to extend the line force description 
(with $\delta$ for ionization; Abbott 1982) and to gain a better understanding of the underlying 
line strength distribution function (Gayley 1995, Puls et al. 2000). 
These analyses have been helpful to our understanding 
of the physics of the line driving force. 
As the line acceleration could simply be fit with two parameters, the 
equation of motion (Eq.~\ref{eq_motion}) was solved in a relatively straightforward manner, delivering simultaneously 
the mass-loss rate (\mdot) and the terminal wind velocity (\vinf). 

\begin{figure}[!ht]
\plotone{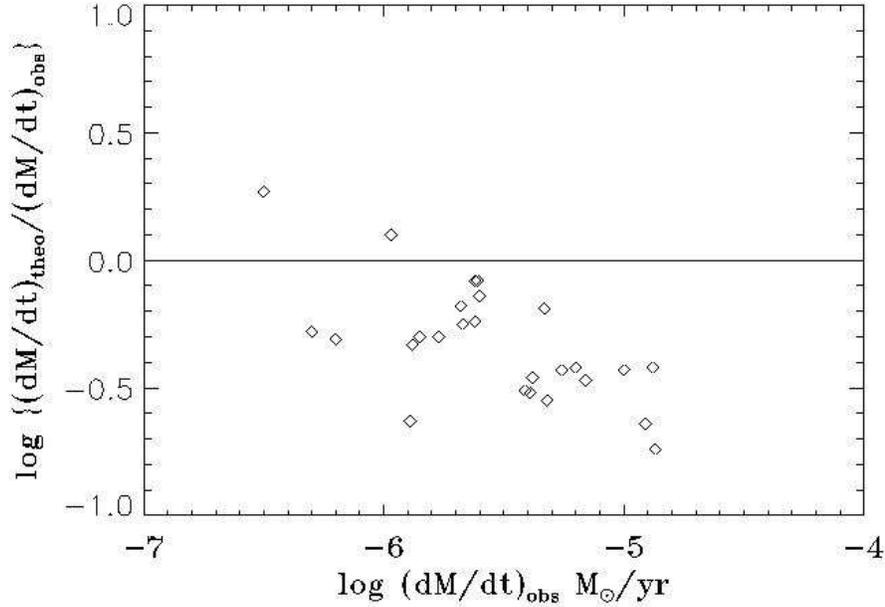}
\caption{A comparison between the modified-CAK mass-loss predictions from the Munich group and 
$H\alpha$ mass-loss rates from Lamers \& Leitherer (1993) -- as a function of the observed mass-loss rate. 
If modified-CAK theory would be in good agreement with observations, the data-points should 
scatter around the solid line, however they show a systematic discrepancy, which grows with increasing 
wind density: the {\it momentum problem}. The figure is taken from Vink (2000).}
\label{f_mom}
\end{figure}

It is important to realize that the CAK predictions yielded mass-loss rates of order $10^{-6}$ $\msunyr$, 
a factor of $\sim$ 100 higher than from the original work by Lucy \& Solomon (1970), and in good accord 
with even today's observations, although we now know {\it a posteriori} that this was somewhat of a 
coincidence, as the contribution of iron (Fe) to the line force greatly exceeds that of carbon for Galactic $Z$. 
Nevertheless, the high mass-loss rates predicted by CAK were pivotal for the realization that
hot-star winds are not an observational curiosity, but that they play a fundamental role in the 
evolution of a massive star. 

Since the time of the original CAK computations, the  
atomic physics has been massively updated in modern codes and occupation numbers have been 
computed in non-LTE by the Munich group (Pauldrach et al. 1990), these modified CAK-type predictions have reached better agreement 
with observational analyses, but have not been able to account for the high-mass loss rates in the 
denser winds of O supergiants. 
This is best exemplified by a comparison between the observed mass-loss 
rates based on the $H\alpha$ equivalent width and modified-CAK predictions from the 
Munich group in Fig.~\ref{f_mom}. 

\begin{figure}[!ht]
\plotfiddle{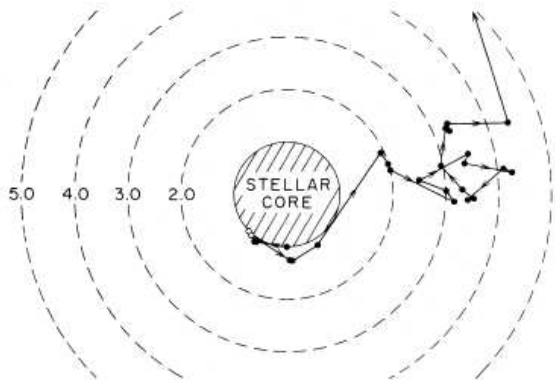}{5.5cm}{0}{57}{58}{-180}{-165}
\caption{The path of a typical Monte Carlo photon that undergoes multiple interactions 
with matter on its way through the stellar wind -- depositing energy and momentum. 
Taken from Abbott \& Lucy (1985). }
\label{f_abbott}
\end{figure}

\subsection{The Monte Carlo method}
It was this momentum problem present in dense O winds as well as Wolf-Rayet (WR) winds that led
Abbott \& Lucy (1985) to develop an entirely new method for predicting the line acceleration via  
Monte Carlo (MC) radiative transfer simulations. Figure~\ref{f_abbott} represents a typical Monte Carlo photon that 
undergoes multiple interactions with the outflowing matter, thereby depositing momentum and energy. 
The philosophy behind the Abbott \& Lucy approach was to {\it adopt} a $\beta$ velocity law 
with an empirically determined \vinf, and derive reliable mass-loss rates from the energy extracted from the radiation field 
-- properly including multiple line scatterings. This resulted in a 
more sophisticated formulation of the line acceleration than in CAK.
In the original work due to Abbott \& Lucy (1985), the ionization calculations were
approximated, but subsequent work on combining the MC approach with non-LTE ionization
computations (Schaerer \& Schmutz 1994, de Koter et al. 1997, Vink et al. 1999) 
have further improved the method. 

The strongest points of the current MC method are the fact that the core-halo approach is 
dropped, the line force is computed for all radii in photosphere and wind, taking 
account of the most extensive line lists (H-Zn), ionization stratification, and multiple 
scatterings on both line and continuum opacity (electron scattering, bound-free and free-free). 
The main drawback is the fact that the derived mass-loss rates are only 
{\it globally} self-consistent, and the momentum equation is usually not solved for, 
i.e. the mass-loss rates are not locally consistent (but see Vink et al. 1999, Vink 2000). 

The basic assumptions that underly all current mass-loss predations are: (i) stationarity, as  
hydrodynamic time-dependent line force computations are not yet feasible; (ii) one-fluid, i.e. 
ion-decoupling is assumed to be negligible; (iii) sphericity, which is probably a rather 
good approximation for stars at solar $Z$ (Harries et al. 1998), but this may no longer hold for sub-solar $Z$; 
(iv) homogeneity, i.e. the wind is assumed to be smooth. 
Mass-loss predictions for clumped winds have yet to be performed.

\section{Results}
Over the last couple of decades the main focus of line-driven wind models concerned 
the Galactic O supergiant $\zeta$ Pup. Comparison between theoretical models 
and observational data highlighted successes and failures of the theory, which 
advanced our knowledge of hot-star winds. 
It is equally important to explore groups of objects, preferentially in different parts 
of the Hertzsprung-Russell Diagram. For this reason, I review below results of wind models 
for OB stars, LBVs and WRs. For mass-loss determinations of hot low-mass stars 
as a function of metallicity, see Pauldrach et al. (1988) and Vink \& Cassisi (2002).

\subsection{OB supergiants at galactic $Z$}
\begin{figure}[!t]
\plotfiddle{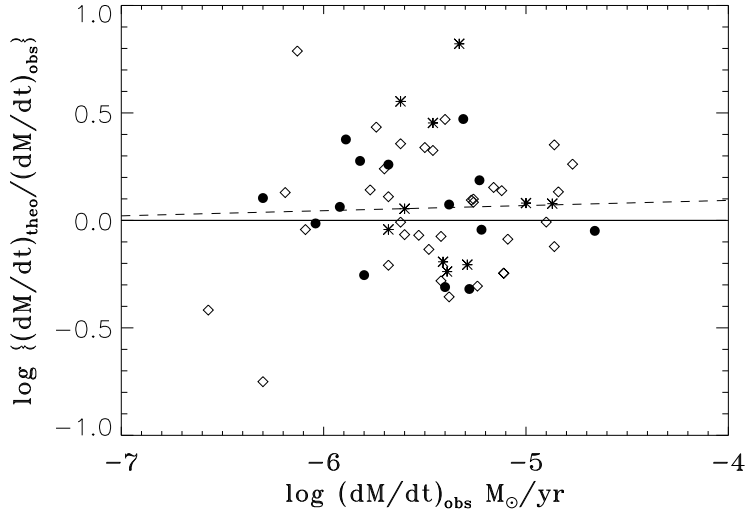}{6.cm}{0}{60}{60}{-190}{-224}
\caption{A comparison between Monte Carlo mass-loss predictions from Vink et al. (2000) and 
a compilation of mass-loss rates -- prior to 2000, as a function of the observed mass-loss rate. 
The comparison shows good average agreement as the scatter is random, i.e. the Monte Carlo method is equally good 
at predicting weak as dense winds. This figure is taken from Vink (2000).}
\label{f_agree}
\end{figure}

The most used mass-loss predictions for OB supergiants at galactic $Z$ 
are those from the Monte Carlo simulations by Vink et al. (2000)\footnote{see www.astro.keele.ac.uk/$\sim$jsv/}, who find the mass-loss 
rate of OB supergiants to roughly scale as:

\begin{equation}
\dot{M}~\propto~L^{2.2}~M^{-1.3}~\Teff^1~(\ratio)^{-1.3}
\label{eq_formula}
\end{equation}
Equation~(\ref{eq_formula}) shows that the mass-loss rate scales strongly 
with luminosity ($L^{2.2}$), much stronger than the scaling of \mdot\ $\propto$ $L^{1.6}$ often quoted in the literature.
The reason is that for the more luminous stars with the denser winds, the MC simulations 
deliver an increasingly larger mass-loss rate than do the modified-CAK predictions. 
As the mass-loss rate {\it also} scales with stellar mass $M^{-1.3}$ (which is often overlooked) and 
when assuming a typical massive star $M-L$ ratio of $L$ 
$\propto$ $M^2$, we find an overall $\mdot$ scaling with $L^{1.6}$ -- in agreement with 
observational scalings (e.g. Howarth \& Prinja 1989). 
The basic success of the Monte Carlo technique is exemplified in Fig.~\ref{f_agree}. It shows that 
properly including multiple scatterings yields equal success for relatively weak 
winds (with log \mdot\ $\sim$ $10^{-7}$ $\msunyr$) as 
for dense winds (with log \mdot\ $\sim$ $10^{-5}$ $\msunyr$). 

We note that for significantly weaker winds, with log $\mdot$ $<$ $10^{-7}$ \msunyr, such as the OVz stars, theory 
and observation appear to show large 
discrepancies (Bouret et al. 2003, Martins et al. 2005).
In fact, even for denser O star winds, large discrepancies have been reported amongst so-called ``reliable'' empirical 
methods (e.g. Fullerton et al. 2005), 
whilst discrepancies have also been reported for B supergiants 
(Vink et al. 2000, Crowther et al. 2005, Trundle \& Lennon 2005). On the other 
hand, good agreement between the Vink et al. mass-loss formulae and a variety of other empirical 
data from radio, $H\alpha$, and UV measurements appears to have been 
achieved (Benaglia et al. 2001, Repolust et al. 2004, Massey et al 2005). It is clear that despite the success 
of resolving the wind momentum problem, there are remaining challenges, in particular with respect to our 
understanding of wind clumping. This may be a vital phenomenon in understanding the source(s) of the 
reported discrepancies. Resolving these discrepancies is obviously pivotal, as the absolute value of 
the mass-loss rate is a crucial input for models of the structure 
and evolution of massive stars.  

\subsection{LBVs and mass loss close to the Eddington limit}
\label{s_edding}
Luminous Blue Variables (LBVs) exhibit growth and shrinking of their stellar radii by factors 
of ten on timescales of about ten years (Humphreys \& Davidson 1994, Weis, these proceedings). 
Leitherer et al. (1994), Vink \& de Koter (2002), and Smith et al. (2004) 
found that LBVs winds are driven by radiation pressure and that the mass-loss variability 
can be attributed to changes in the temperature and ionization of the dominant Fe ions. For 
an explanation of these mass loss changes at these ``bi-stability jumps'', I 
refer the reader to Vink et al. (1999). 

\begin{figure}[!t]
\plotfiddle{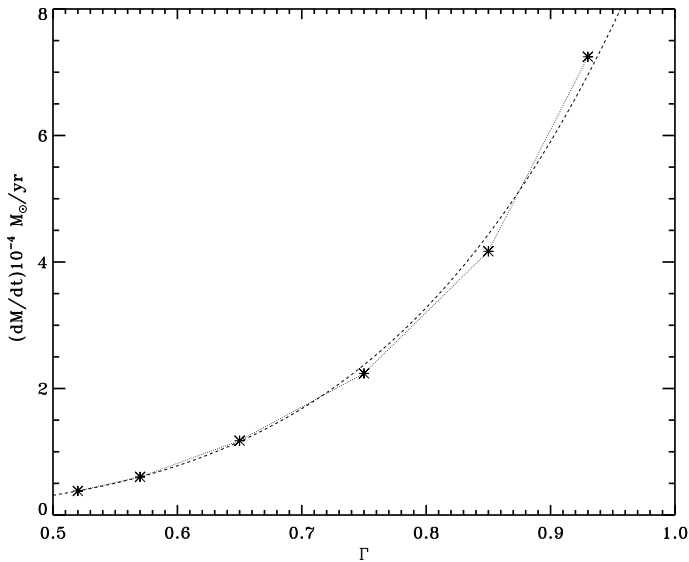}{6.cm}{0}{40}{40}{-157}{-34}
\vspace*{0.8cm}
\caption{The mass-loss rate of VMS (100 - 300 $\msun$) as a function of $\Gamma$ (in the range 0.5 -- 0.9). The dotted line 
shows that \mdot\ is predicted to increase drastically when approaching the Eddington limit. 
The dashed line represents \mdot\ $\propto$ $\Gamma^5$.}
\label{f_gamma}
\end{figure}

Vink \& de Koter (2002) found that for LBVs (with $\Gamma$ ($\kappa_{\rm e}L/4\pi c g M$) $\sim$ 0.5) the mass-loss rate 
is strongly dependent on the stellar mass, \mdot\ $\propto$ 
$M^{1.8}$, a much stronger dependence as for OB supergiants that show \mdot\ $\propto$ 
$M^{1.3}$. 

We may wonder what this implies for objects that find themselves in even closer proximity 
to the Eddington limit. This is 
not only relevant for the unstable evolutionary phases of ``normal'' massive stars, but especially 
relevant for VMS already from the very beginning of their life.

For this reason, we present preliminary MC mass-loss calculations for VMS in 
the range 100 -- 300 $\msun$, with log$(L/\lsun)$ 6.3 -- 7.0 (Kudritzki 2002) 
and $\Gamma$ in the range of 0.5 -- 0.9. The dotted line in Fig.~\ref{f_gamma} shows that mass loss 
increases dramatically when approaching the Eddington limit. 
For comparative purposes, the dashed line represents the function 

\begin{equation}
\dot{M}~\propto~\Gamma^5
\label{eq_gamma}
\end{equation}

\subsection{OB winds as a function of $Z$}
Given that radiation pressure on metal lines has been identified to be the driving mechanism 
for O stars, it follows rather straightforwardly that \mdot\ is smaller at lower $Z$. 
Therefore, the question is not so much {\it if} O star winds are $Z$-dependent, but 
rather {\it by how much}? 

Modified CAK predictions over the last decades have predicted 
that \mdot\ $\propto$ $Z^{m}$, where $m$ has been found to be between 1/2 and 1 (Abbott 1982, Kudritzki et al. 1987). 
The first order effect of this scaling is that
the CAK $k$ parameter from Eq.~(\ref{eq_cak}) -- representing the sheer number of lines -- is 
strongly $Z$-dependent.
Second order effects, relating to how $\alpha$ varies with $Z$, have a particularly strong 
effect on \vinf. For a full discussion of how the CAK parameters are expected to vary with $Z$, see Puls et al. (2000).

\begin{figure}[!ht]
\plotfiddle{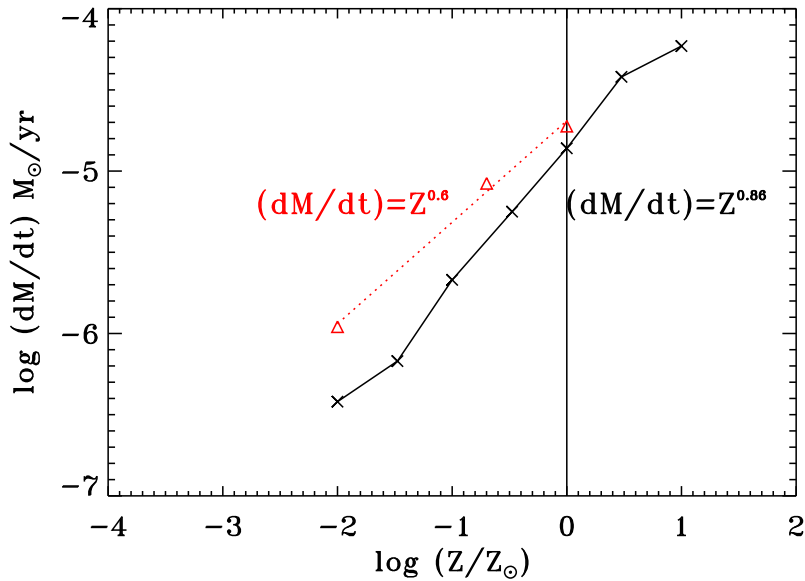}{6.5cm}{0}{68}{68}{-217}{-264}
\caption{Mass-loss predictions as a function of $Z$. The solid line indicates the $\mdot(Z)$ dependence from 
MC simulations by Vink et al. (2001). The dotted line is for mass-loss predictions with an updated version of 
the modified CAK theory due to Kudritzki (2002) for stars in the observable Universe (log$(Z/\zsun) \ga 1/100$), for objects with similar 
stellar parameters. See text for a discussion on the difference in slopes.}
\label{f_kudrcomp}
\end{figure}

Over the years, most stellar evolution models have relied on a square-root dependence 
of how \mdot\ scales with $Z$, i.e. $m$ = 0.5 (Kudritzki et al. 1987), however I argue that the 
real dependence is likely to be closer to a linear dependence, $m$ $\ge$ 0.7. 
The difference in the two line-force approaches and the resulting implications for the metallicity slopes are 
illustrated in Fig.~\ref{f_kudrcomp} by comparing the modified CAK predictions 
of Kudritzki (2002), where depth-dependent force multiplier parameters are employed, with 
the MC predictions of Vink et al. (2001).

As can be seen in Fig.~\ref{f_kudrcomp}, the Vink et al. (2001) predictions show a steeper slope ($m$ = 0.85) 
of the $Z$-dependence than do the Kudritzki (2002) predictions (where $m$ = 0.5 - 0.6). 
The reasons for this are thought to be the following. CAK-type wind models employ 
a single line approach, which means that unattenuated stellar flux is offered to each line. 
This may result in overestimating the mass-loss rate, unless the winds are dense enough so that multiple scatterings 
are important, and the mass-loss rate is underestimated instead. 
The single line approach therefore probably leads to a systematic discrepancy between theory and observations, in that  
the $\mdot-L$ dependence is too weak, and the $\mdot$-$Z$ dependence is not steep enough. 

On the other hand, most current MC simulations do not solve the momentum equation, and therefore 
do not account for the dependence of \vinf\ with $Z$. This will likely lead to an overestimate of $m$, and one should therefore lower 
the $m$ value for O supergiants from 0.85 to 0.7. For these reasons, the use of $m$ = 0.7 
as the ``minimum'' value for the mass-loss scaling with $Z$ for O stars is advisable 
(Vink et al. 2001). 

\subsection{Wolf-Rayet winds as a function of $Z$}
\label{s_wr}
Unlike the case of OB supergiants, the question of a WR mass-loss versus $Z$ dependence 
was whether {\it WR winds scale with $Z$ at all}. This  
despite empirical indications that WR winds may well depend on the metal content of the host 
galaxy (Crowther et al. 2002, Hadfield et al. 2005).

Even when it became rather well-established that WR winds are radiatively-driven (Lucy \& Abbott 1993, 
Gayley et al. 1995, Nugis \& Lamers 2002, Hillier 2003, Gr\"afener et al. 2005), 
the question of a $Z$-dependence remained highly controversial, with some evolutionary modellers 
adopting a square-root extrapolation from modified-CAK O-star results, whereas others assumed 
WR winds not to weaken at lower environmental $Z$. 

\begin{figure}[!ht]
\plotfiddle{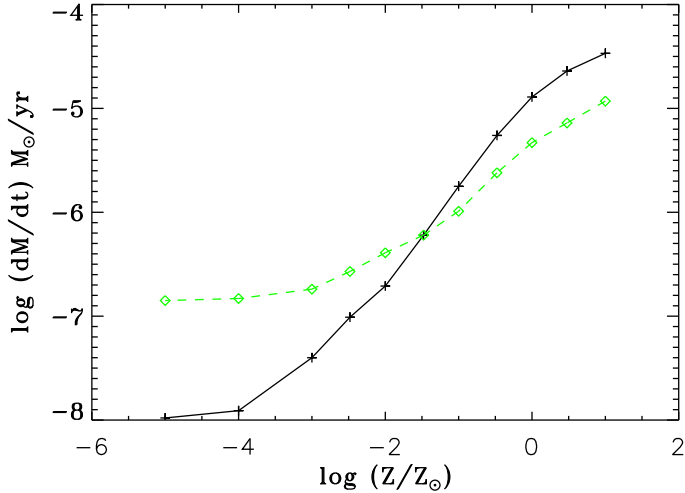}{5.5cm}{0}{58}{58}{-174}{-224}
\caption{WR mass-loss predictions as a function of $Z$. 
The dark line indicates the $\mdot(Z)$ dependence for 
late-type WN stars, whereas the lighter line 
represents late-type WC star predictions. The slope of 
the WN models is very similar to OB stars with $m$ = 0.85. 
The slope for WC stars is much smaller, and flattens off 
below $(Z/\zsun)$ $=$ $10^{-3}$.}
\label{f_wr}
\end{figure}

The reason often put forward that WR winds might not be $Z$-dependent is that WR stars {\it enrich themselves} by 
burning helium into carbon in their cores. It may be this self-enriched carbon 
material that would primarily be responsible for the radiative driving, whilst the trace amounts 
of Fe responsible for driving the OB star winds, would not be important and it should not matter 
in which galaxy a WR star would reside. 

To answer the question of whether WR winds are $Z$-dependent, Vink \& de Koter (2005) performed 
WR mass-loss predictions at a range of $Z$, unravelling which elements drive 
the winds of the WR subtypes, and whether the same chemical species dominate the line force for 
all $Z$. Their results are shown in Fig.~\ref{f_wr}. The dark line denotes the $\mdot(Z)$ dependence for 
late-type WN stars. The slope is identical to that of OB stars: $m$ = 0.85, as WN winds are dominated  
by Fe, and the increase of secondary nitrogen in the 
photosphere is easily offset by the decrease in carbon and oxygen. 
These initial results strongly suggest that WN winds should be 
scaled with $Z$, with important implications for the final evolutionary phases 
of massive stars at low $Z$, such as for black hole formation, supernovae and the determination of the 
threshold metallicity to form a GRB (Woosley \& Heger 2006). 

The lighter line in Fig.~\ref{f_wr} represents late-type WC wind predictions: the 
slope is smaller c.f. OB/WN stars.
However, it flattens off once the metal content drops below $(Z/\zsun)$ $\le$ $10^{-3}$, as the ratio of 
carbon over Fe to the line force increases at lower $Z$. 
At solar $Z$, the large number of Fe lines overwhelms the carbon contribution to the line force near the 
photosphere, but as $Z$ drops these Fe lines no longer reach the necessary strength (Sobolev optical depth) 
to be able to contribute to the line force, and carbon takes over.

\section{Cosmological implications and future work}
It is often commented that mass loss at extremely low ($Z$ $<$ $10^{-3}$)  
is negligible because of the rapidly decreasing role of metal-line 
driving once $Z$ approaches zero. These assertions are based on modified CAK predictions 
of O stars at low $Z$ (Kudritzki 2002, Krti\v{c}ka, these proceedings). Although 
it must be true that the line-driving on Fe lines decreases in ``normal'' 
($\Gamma$ $\le$ 0.5) massive stars that we see in galaxies today, 
I have highlighted two pieces of the jigsaw that are often 
overlooked: (i) massive stars in close proximity to the Eddington limit (Sect.~\ref{s_edding}), and (ii) 
the role of radiation pressure due to carbon via self-enrichment (Sect.~\ref{s_wr}).

Recent numerical simulations on the formation of Population {\sc iii} 
indicate that they were both massive and luminous and may well have been in 
close proximity to the Eddington limit. In addition, there 
are reasons to believe their rotational surface speeds were larger (e.g. Meynet \& Maeder 2004) 
which may have caused additional mass loss close to the Eddington-Omega limit (Langer 1998). 
However, this is far from the entire story, as rotation also induces mixing. 
Due to rotationally-induced mixing of primary nitrogen and 
carbon (e.g. Meynet et al. 2005, Yoon \& Langer 2006) into the atmospheres of 
zero-metallicity massive stars, the flattening in Fig.~\ref{f_wr} suggests that carbon driving may 
have significantly boosted the mass-loss rates of the first stars, 
thereby potentially preventing the occurrence of pair-instability supernovae that 
are believed to occur for initial masses in the range $M$ = 140 -- 260 $\msun$ (Heger et al. 2003), if mass 
loss from the first stars were negligible.

To start answering these cosmologically important questions, there remains a number of wind 
aspects to be resolved. As I have mentioned, the MC predictions have 
achieved a large number of successes, but to be able to confidently predict the mass-loss rates 
of the first luminous objects, where one cannot rely on direct empirical constraints, a solution of 
the wind momentum equation is warranted. Furthermore, we need to understand the physics of wind clumping. Hydrodynamical 
work on wind instabilities and distance-dependent clumping is well 
underway (e.g. Runacres \& Owocki 2002), whereas empirical constraints may be obtained from polarimetric 
monitoring (see Moffat \& Robert 1994 for WR stars, and Nordsieck et al. 2001 and 
Davies et al. 2005 for the even more strongly clumped winds of LBVs). Once we are able to predict 
the mass-loss rates and terminal velocities of structured winds from first principles, we can  
address the questions as to the quantitative role of the winds of the first stars, and whether they could  
have enriched the IGM -- even before the first supernova.

\acknowledgements 
I would like to thank the SOC for inviting me to give this review, and Alex 
de Koter and Henny Lamers for fruitful collaboration.


\end{document}